\documentclass[12pt]{article}

\usepackage{graphicx}
\usepackage{float}
\usepackage{cite}
\usepackage{amsfonts}
\usepackage{amssymb}
\usepackage{amsmath}

\def\gtwid{\mathrel{\raise.3ex\hbox{$>$\kern-.75em\lower1ex\hbox{$\sim$}}}}
\def\ltwid{\mathrel{\raise.3ex\hbox{$<$\kern-.75em\lower1ex\hbox{$\sim$}}}}
\def\square{\kern1pt\vbox{\hrule height 1.2pt\hbox{\vrule width 1.2pt\hskip 3pt
   \vbox{\vskip 6pt}\hskip 3pt\vrule width 0.6pt}\hrule height 0.6pt}\kern1pt}

\begin{document}

\begin{titlepage}

\begin{flushright}
UFIFT-QG-21-05
\end{flushright}

\vskip 0.2cm

\begin{center}
{\bf Large Logarithms from Quantum Gravitational Corrections to a
Massless, Minimally Coupled Scalar on de Sitter}
\end{center}

\vskip 0.2cm

\begin{center}
D. Glavan$^{1*}$, S. P. Miao$^{2\star}$, T. Prokopec$^{3\dagger}$ and
R. P. Woodard$^{4\ddagger}$
\end{center}
\vspace{0.cm}
\begin{center}
\it{$^{1}$ CEICO, Institute of Physics of the Czech Academy of Sciences (FZU), \\
Na Slovance 1999/2, 182 21 Prague 8, CZECH REPUBLIC}
\end{center}
\vspace{-0.3cm}
\begin{center}
\it{$^{2}$ Department of Physics, National Cheng Kung University, \\
No. 1 University Road, Tainan City 70101, TAIWAN}
\end{center}
\vspace{-0.3cm}
\begin{center}
\it{$^{3}$ Institute for Theoretical Physics, Spinoza Institute \& EMME$\Phi$, \\
Utrecht University, Postbus 80.195, 3508 TD Utrecht, THE NETHERLANDS}
\end{center}
\vspace{-0.3cm}
\begin{center}
\it{$^{4}$ Department of Physics, University of Florida,\\
Gainesville, FL 32611, UNITED STATES}
\end{center}

\vspace{0.2cm}

\begin{center}
ABSTRACT
\end{center}
We consider single graviton loop corrections to the effective 
field equation of a massless, minimally coupled scalar on de
Sitter background in the simplest gauge. We find a large 
temporal logarithm in the approach to freeze-in at late times,
but no correction to the feeze-in amplitude. We also find a 
large spatial logarithm (at large distances) in the scalar 
potential generated by a point source, which can be explained 
using the renormalization group with one of the higher 
derivative counterterms regarded as a curvature-dependent field 
strength renormalization.  We discuss how these results set the 
stage for a project to purge gauge dependence by including 
quantum gravitational corrections to the source which disturbs 
the effective field and to the observer who measures it.

\begin{flushleft}
PACS numbers: 04.50.Kd, 95.35.+d, 98.62.-g
\end{flushleft}

\vspace{0.2cm}

\begin{flushleft}
$^{*}$ e-mail:glavan@fzu.cz \\
$^{\star}$ e-mail: spmiao5@mail.ncku.edu.tw \\
$^{\dagger}$ e-mail: T.Prokopec@uu.nl \\
$^{\ddagger}$ e-mail: woodard@phys.ufl.edu
\end{flushleft}

\end{titlepage}

\section{Introduction}

One of the most profound predictions of primordial inflation is that the 
accelerated expansion literally rips long wavelength quanta out of the vacuum 
\cite{Starobinsky:1979ty}. This is what produced the tensor power spectrum
$\Delta_{h}^2(k)$ \cite{Starobinsky:1985ww}. The occupation number $N(\eta,k)$
of a single polarization of wave number $\vec{k}$ at conformal time $\eta$ is 
simply staggering,
\begin{equation}
N(\eta,k) = \frac{\pi \Delta_{h}^2(k)}{64 G k^2} \!\times\! a^2(\eta) \; ,
\label{occunum}
\end{equation}
where $G$ is Newton's constant and $a(\eta)$ is the scale factor, which we 
remind the reader grows exponentially rapidly in co-moving time. 

The tensor power spectrum is the primary, tree order signal of the production 
of inflationary gravitons. However, there must be secondary, loop effects from
the interactions of these gravitons with each other and with other particles.
Among these effects are: 
\begin{enumerate}
\item{The self-gravitation between inflationary gravitons may slow the 
expansion rate \cite{Tsamis:1996qq,Tsamis:1996qm};}
\item{Inflationary gravitons correct the linearized Einstein equation
\cite{Tsamis:1996qk,Tan:2021ibs}, which enhances the field strength of 
gravitational radiation \cite{Mora:2013ypa,Tan:2021lza} and has the potential 
to change the force of gravity \cite{Park:2015kua};}
\item{Inflationary gravitons correct the linearized Dirac equation
\cite{Miao:2005am,Miao:2012bj}, which enhances the field strength of fermions
\cite{Miao:2006gj};}
\item{Inflationary gravitons correct the field equations for a massless,
minimally coupled scalar \cite{Kahya:2007cm}, but make no significant 
change in the field strength of scalar radiation \cite{Kahya:2007bc};}
\item{Inflationary gravitons correct Maxwell's equation \cite{Leonard:2013xsa},
which enhances the field strength of electromagnetic radiation 
\cite{Wang:2014tza} and makes significant changes to the response to charges 
and currents at large distances and late times \cite{Glavan:2013jca}; and}
\item{Inflationary gravitons correct the field equation for a massless,
conformally coupled scalar \cite{Boran:2014xpa,Boran:2017fsx,Glavan:2020gal},
but do not make significant changes, either in the propagation of dynamical
scalars or in the scalar exchange potential \cite{Glavan:2020ccz}.}
\end{enumerate}

No one doubts that a classical ensemble of gravitational radiation would 
change kinematics and forces; this is the basis for the proposal to detect
gravitational radiation using the timing of pulsars \cite{Detweiler:1979wn,
Lorimer:2008se}. However, graviton propagators do require gauge fixing, and
it has been argued that the apparent effects of inflationary gravitons are 
artifacts of the gauge \cite{Garriga:2007zk,Tsamis:2007is,Higuchi:2011vw,
Miao:2011ng,Morrison:2013rqa,Miao:2013isa,Frob:2014fqa,Woodard:2015kqa}.
These doubts persist in spite of the fact that similar effects derive from 
loops of massless, minimally coupled scalars \cite{Park:2011ww,Park:2015kua}, 
which experience the same growth (\ref{occunum}) as inflationary gravitons 
and require no gauge fixing.  

\begin{figure}[ht]
\hspace{1cm} \includegraphics[width=11.0cm,height=0.5cm]{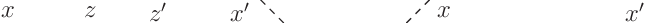}
\caption{\footnotesize The left diagram shows how the self-mass 
(\ref{M0}) contributes to massive scalar scattering. The diagram 
on the right gives the contribution from graviton correlations 
between the vertices. Solid lines represent the massless scalar, 
wavy lines represent the graviton, and dashed lines are massive 
scalars.}
\label{BB84}
\end{figure}

A technique has been developed for removing gauge dependence from
effective field equations by including quantum gravitational correlations
with the source which disturbs the effective field and the observer who
detects it \cite{Miao:2017feh}. The procedure is to build the same diagrams
(Figure~\ref{BB84} gives two examples) that would go into an S-matrix 
element, and then simplify them with a series of relations derived by 
Donoghue \cite{Donoghue:1993eb,Donoghue:1994dn,Donoghue:1996mt} to capture 
the infrared physics. In the end only vestigial traces of the source and 
observer remain, and each of the simplified diagrams can be regarded as a 
correction to the 1PI 2-point function in the linearized, effective field 
equation.
\vskip 1.5cm

It is worthwhile describing this for a massless, minimally coupled scalar 
on flat space background ($g_{\mu\nu} \equiv \eta_{\mu\nu} + \sqrt{16 \pi G}
\, h_{\mu\nu}$) in the 2-parameter family of covariant gauge fixing functions,
\begin{equation}
\mathcal{L}_{GF} = -\frac1{2 \alpha} \eta^{\mu\nu} F_{\mu} F_{\nu} \qquad , 
\qquad F_{\mu} = \eta^{\rho\sigma} \Bigl( h_{\mu \rho , \sigma} - 
\frac{\beta}{2} h_{\rho\sigma , \mu} \Bigr) \; . \label{gauge}
\end{equation}
The renormalized self-mass is \cite{Miao:2017feh},
\begin{eqnarray}
-i M^2(x;x') & = & \mathcal{C}_0(\alpha,\beta) \times 
\frac{G \partial^6}{4 \pi^3} \Biggl[ \frac{ \ln(\mu^2 \Delta x^2)}{\Delta x^2} 
\Biggr] \quad , \quad \Delta x^2 \equiv (x \!-\! x')^2 \; , \label{M0} 
\qquad \\
\mathcal{C}_0(\alpha,\beta) & = & +\frac34 -\frac34 \times \alpha - 
\frac32 \times \frac1{\beta \!-\! 2} + \frac34 \times 
\frac{(\alpha \!-\! 3)}{(\beta \!-\! 2)^2} \; . \label{C0}
\end{eqnarray}
Now imagine quantum gravitational corrections to the scattering of 
two massive scalars by the exchange of such a massless scalar. 
Figure~\ref{BB84} shows two of the many diagrams which contribute. After
applying the Donoghue Identities, each of these contributions can be regarded
as a correction to the self-mass, having the same spacetime dependence as
(\ref{M0}) but multiplied by different gauge dependent coefficients. 
Table~\ref{Cab} lists each contribution, and one can see that the sum is
indeed independent of $\alpha$ and $\beta$.
\begin{table}[H]
\setlength{\tabcolsep}{8pt}
\def\arraystretch{1.5}
\centering
\begin{tabular}{|@{\hskip 1mm }c@{\hskip 1mm }||c|c|c|c|c|}
\hline
$i$ & $1$ & $\alpha$ & $\frac1{\beta-2}$ & $\frac{(\alpha-3)}{(\beta-2)^2}$ &
{\rm Description} \\
\hline\hline
0 & $+\frac34$ & $-\frac34$ & $-\frac32$ & $+\frac34$ &
{\rm scalar\ exchange} \\
\hline
1 & $0$ & $0$ & $0$ & $+1$ & {\rm vertex-vertex} \\
\hline
2 & $0$ & $0$ & $0$ & $0$ & {\rm vertex-source,observer} \\
\hline
3 & $0$ & $0$ & $+3$ & $-2$ & {\rm vertex-scalar} \\
\hline
4 & $+\frac{17}4$ & $-\frac34$ & $0$ & $-\frac14$ &
{\rm source-observer} \\
\hline
5 & $-2$ & $+\frac32$ & $-\frac32$ & $+\frac12$ &
{\rm scalar-source,observer} \\
\hline\hline
Total & $+3$ & $0$ & $0$ & $0$ & \\
\hline
\end{tabular}
\caption{\footnotesize The gauge dependent factors $C_i(\alpha,\beta)$ 
for each contribution to the invariant scalar self-mass-squared, and their
gauge-independent sum. Figure~\ref{BB84} shows the $i=0$ and $i=1$ diagrams.}
\label{Cab}
\end{table}

In position space all diagrams consist of products of (possibly 
differentiated) massive and massless propagators (from the scalar and
the graviton), $i\Delta_m(x;x')$ and $i\Delta(x;x')$, respectively. 
All the 3-point and 4-point diagrams can be reduced to 2-point form by 
applying the Donoghue Identities \cite{Miao:2017feh},
\begin{eqnarray}
\lefteqn{i\Delta_m(x;y) i\Delta(x;x') i\Delta(y;x') \longrightarrow
\frac{i\delta^D(x \!-\! y)}{2 m^2} \Bigl[ i\Delta(x;x')\Bigr]^2 \; ,}
\label{IntID1} \\
\lefteqn{\partial^{x}_{\mu} i\Delta(x;x') \partial_{y}^{\mu} i\Delta(y;y') 
i\Delta_m(x;y) i\Delta_m(x';y') } \nonumber \\
& & \hspace{4.5cm} \longrightarrow \frac{i\delta^D(x \!-\!y) 
i\delta^D(x' \!-\! y')}{2 m^2} \Bigl[ i\Delta(x;x')\Bigr]^2
\; , \qquad \label{IntID2} \\
\lefteqn{ \partial^{x}_{\mu}  i\Delta(x;y') \partial^{\mu}_{y} i\Delta(y;x') 
i\Delta_m(x;y) i\Delta_m(x';y') } \nonumber \\
& & \hspace{4.5cm} \longrightarrow -\frac{i\delta^D(x \!-\! y) 
i\delta^D(x' \!-\! y')}{2 m^2} \Bigl[ i\Delta(x;x')\Bigr]^2
\; . \qquad \label{IntID3}
\end{eqnarray}
Any 2-point contribution so obtained can be regarded as a correction 
to the self-mass through a trivial identity based the massless propagator 
equation $\partial^2 i\Delta(x;x') = i\delta^D(x - x')$,
\begin{equation}
f(x;x') = -\int d^Dz \, i\Delta(x;z) \int d^Dz' \, i\Delta(x';z') \times
\partial_z^2 \partial_{z'}^2 f(z;z') \; . \label{keytrick}
\end{equation}

The procedure just described has been implemented on flat space 
background for scalars \cite{Miao:2017feh} and for electromagnetism 
\cite{Katuwal:2020rkv}. Generalizing it to de Sitter will be challenging 
because the Hubble parameter $H$ permits more varied spacetime dependence 
than (\ref{M0}) on the dimensionless product $H^2 \Delta x^2$. Our program 
is therefore to find the simplest venue for implementing the gauge purge 
on de Sitter background, and then check gauge independence using the 
family of de Sitter breaking gauges analogous to (\ref{gauge})
\cite{Glavan:2019msf}. In addition to simplicity, we require a system for 
which the potentially gauge dependent computation shows big effects, 
because there is little point to removing gauge dependence from a small 
or null effect. Previous studies have revealed that graviton corrections 
to massless, conformally coupled scalars are simple but do not engender 
significant effects \cite{Glavan:2020gal,Glavan:2020ccz}. In this paper 
we show that the massless, minimally coupled scalar provides the system 
we seek.

In section 2 of this paper we compute the single graviton loop contribution
to the self-mass $-i M^2(x;x')$ of a massless, minimally coupled scalar on
de Sitter background. Section 3 uses this result to quantum-correct the
linearized effective scalar field equation. Solving this equation reveals
no significant 1-loop correction to the field strength of scalar radiation,
but a large logarithmic correction to the scalar exchange potential. We 
also explain the large logarithm using a version of the renormalization 
group. Our conclusions comprise section 4.

\section{Graviton Loop Contribution to $-i M^2(x;x')$}

The purpose of this section is to compute the 1-graviton loop contribution
to the 1PI (one-particle-irreducible) 2-point function of a massless,
minimally coupled scalar on de Sitter background. We begin by precisely
defining $-i M^2(x;x')$, analytically and diagrammatically, and by giving 
the required Feynman rules. We next employ dimensional regularization to 
evaluate first the simplest diagram and then the more complicated one. The 
section closes with a discussion of renormalization.

\subsection{Feynman Rules}

The bare Lagrangian in $D$ spacetime dimensions is,
\begin{equation}
\mathcal{L} = \frac{[R \!-\! (D\!-\!2) \Lambda] \sqrt{-g}}{16 \pi G} 
-\frac12 \partial_{\mu} \phi \partial_{\nu} \phi g^{\mu\nu} \sqrt{-g} 
\qquad , \qquad \Lambda \equiv (D\!-\!1) H^2 \; , \label{MMCSLag}
\end{equation}
where $G$ is Newton's constant and $\Lambda$ is the cosmological constant.
We define the graviton field $h_{\mu\nu}(x)$ as a perturbation of the 
conformally rescaled metric,
\begin{equation}
g_{\mu\nu}(x) \equiv a^2(\eta) \widetilde{g}_{\mu\nu}(x) \equiv a^2(\eta) 
\Bigl[ \eta_{\mu\nu} + \kappa h_{\mu\nu}\Bigr] \qquad , \qquad a(\eta) 
\equiv -\frac1{ H \eta} \; , \label{graviton}
\end{equation}
where $\kappa^2 \equiv 16 \pi G$ is the loop-counting parameter and 
$\eta < 0$. Our signature is spacelike, and we employ an overlined tensor 
to denote the suppression of its temporal components,
\begin{equation}
\overline{\eta}_{\mu\nu} \equiv \eta_{\mu\nu} + \delta^0_{~\mu} 
\delta^0_{~\nu} \qquad , \qquad \overline{\partial}_{\mu} \equiv 
\partial_{\mu} - \delta^0_{~\mu} \partial_0 \; . \label{bardef}
\end{equation}

\begin{figure}[H]
\hspace{1cm} \includegraphics{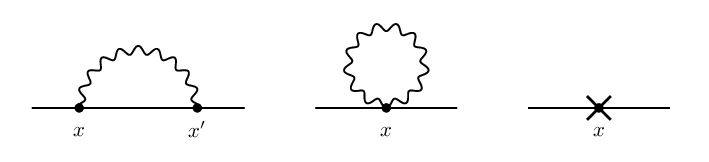}
\caption{\footnotesize 1-graviton loop contributions to $-i M^2(x;x')$
corresponding to the three terms of expression (\ref{MsqVEV}). Graviton 
lines are wavy, scalar lines are straight and counterterms are denoted 
by a cross.}
\label{scalar}
\end{figure}

The 1-graviton loop contribution to the scalar self-mass can be represented
as the in-out matrix element of variations of the action $S[g,\phi]$ and the 
counterterm action $\Delta S[g,\phi]$ (discussed in subsection 2.4),
\begin{eqnarray}
\lefteqn{ -i M^2(x;x') = \Bigl\langle \Omega^{\rm out} \Bigl\vert T^*\Biggl\{ 
\Bigl[\frac{i \delta S[g,\phi]}{\delta \phi(x)}\Bigr]_{h \phi} \!\times\!
\Bigl[\frac{i \delta S[g,\phi]}{\delta \phi(x')}\Bigr]_{h \phi} } 
\nonumber \\
& & \hspace{4.2cm} + \Bigl[\frac{i \delta^2 S[g,\phi]}{\delta \phi(x) 
\delta \phi(x')}\Bigr]_{h h} + \Bigl[\frac{i \delta^2 \Delta S[g,\phi]}{
\delta \phi(x) \delta \phi(x')}\Bigr]_{1} \Biggr\} \Bigr\vert \Omega^{\rm in}
\Bigr\rangle . \qquad \label{MsqVEV}
\end{eqnarray}
The $T^*$-ordering symbol in expression (\ref{MsqVEV}) indicates that 
derivatives are taken outside time-ordering; the subscripts of 
square-bracketed variations indicate how many perturbative fields contribute 
to the 1-loop result. The associated diagrams are shown in 
Figure~\ref{scalar}.

The propagators mostly depend on the de Sitter length function $y(x;x')$,
\begin{equation}
y(x;x') \equiv a a' H^2 \Delta x^2(x;x') \equiv a a' H^2 \Bigl[ 
\Bigl\Vert \vec{x} \!-\! \vec{x}' \Bigr\Vert^2 - \Bigl( \vert \eta \!-\! 
\eta'\vert \!-\! i \epsilon\Bigr)^2 \Bigr] \; . \label{ydef}
\end{equation}
The scalar propagator is \cite{Onemli:2002hr,Onemli:2004mb},
\begin{equation}
i\Delta_A(x;x') = A(y) + k \ln(a a') \qquad , \qquad k \equiv
\frac{H^{D-2}}{(4 \pi)^{\frac{D}2}} \frac{\Gamma(D \!-\! 1)}{
\Gamma(\frac{D}2)} \; , \label{DeltaA}
\end{equation}
where the derivative of $A(y)$ is,
\begin{eqnarray}
\lefteqn{A'(y) = -\frac{H^{D-2}}{4 (4 \pi)^{\frac{D}2}} \Biggl\{ 
\Gamma\Bigl(\frac{D}2\Bigr) \Bigl( \frac{4}{y}\Bigr)^{\frac{D}2} + 
\Gamma\Bigl( \frac{D}2 \!+\! 1\Bigr) \Bigl( \frac{4}{y}\Bigr)^{
\frac{D}2 - 1} } \nonumber \\
& & \hspace{2cm} + \sum_{n=0}^{\infty} \Biggl[ \frac{\Gamma(n \!+\! 
\frac{D}2 \!+\! 2)}{\Gamma(n \!+\! 3)} \Bigl( \frac{y}{4}\Bigr)^{
n-\frac{D}2 + 2} - \frac{\Gamma(n \!+\! D)}{\Gamma(n \!+\! \frac{D}2 
\!+\! 1)} \Bigl( \frac{y}{4}\Bigr)^{n} \Biggr] \Biggr\} . \qquad 
\label{Aprime}
\end{eqnarray}
The undifferentiated result can be inferred from the coincidence limit,
\begin{equation}
i\Delta_A(x;x) = k \Bigl[-\pi {\rm cot}\Bigl( \frac{D \pi}{2}\Bigr) 
+ 2 \ln(a) \Bigr] \; . \label{coincprop}
\end{equation}

Our gauge fixing term is a de Sitter breaking analog of (\ref{gauge})
for $\alpha = \beta = 1$ \cite{Tsamis:1992xa,Woodard:2004ut},
\begin{equation}
\mathcal{L}_{GF} = -\frac12 a^{D-2} \eta^{\mu\nu} F_{\mu} F_{\nu}
\;\; , \;\; F_{\mu} = \eta^{\rho\sigma} \Bigl[h_{\mu\rho , \sigma}
\!-\! \frac12 h_{\rho\sigma , \mu} \!+\! (D \!-\! 2) a H h_{\mu\rho}
\delta^0_{~\sigma}\Bigr] . \label{dSgauge}
\end{equation}
In this gauge the graviton propagator is the sum of three constant
tensor factors times scalar propagators,
\begin{equation}
i \Bigl[\mbox{}_{\mu\nu} \Delta_{\rho\sigma}\Bigr](x;x') = 
\sum_{I=A,B,C} \Bigl[\mbox{}_{\mu\nu} T^I_{\rho\sigma}\Bigr]
\!\times\! i\Delta_I(x;x') \; . \label{gravprop}
\end{equation}
The $A$-type propagator is the same as the scalar propagator
(\ref{DeltaA}). The $B$-type and $C$-type propagators are for minimally
coupled scalars with masses $M^2_B = (D-2) H^2$ and $M^2_C = 2 (D-3) H^2$,
which can be expressed as,
\begin{equation}
i\Delta_B = -\frac{[(4 y \!-\! y^2) A'(y) \!+\! (2 \!-\! y) k]}{2 (D\!-\!2)} 
\quad , \quad i\Delta_C = \frac{(2 \!-\! y) i\Delta_B}{2} + 
\frac{k}{D \!-\! 3} \; . \label{DeltaBC}
\end{equation}
And the constant tensor factors $[\mbox{}_{\mu\nu} T^I_{\rho\sigma}]$ are,
\begin{eqnarray} 
\Bigl[\mbox{}_{\mu\nu} T^A_{\rho\sigma}\Bigr] = 2 \overline{\eta}_{\mu (\rho}
\overline{\eta}_{\sigma) \nu} - \frac{2}{D\!-\!3} \overline{\eta}_{\mu\nu}
\overline{\eta}_{\rho\sigma} \quad , \quad \Bigl[\mbox{}_{\mu\nu} T^B_{\rho\sigma}
\Bigr] = -4 \delta^0_{~(\mu} \overline{\eta}_{\nu) (\rho} \delta^0_{~ \sigma)} 
\; , \qquad \label{TATB} \\
\Bigl[\mbox{}_{\mu\nu} T^C_{\rho\sigma}\Bigr] = \frac{2 E_{\mu\nu} E_{\rho\sigma}}{
(D\!-\!2) (D\!-\!3)} \quad , \quad E_{\mu\nu} \equiv (D\!-\!3) \delta^0_{~\mu}
\delta^0_{~\nu} + \overline{\eta}_{\mu\nu} \; . \qquad \label{TCE} 
\end{eqnarray}
Here and henceforth, parenthesized indices are symmetrized.

A hatted 2nd rank tensor denotes the trace-reversal,
\begin{equation}
\widehat{h}_{\mu\nu} \equiv h_{\mu\nu} - \frac12 \eta_{\mu\nu} h \; .
\end{equation}
Trace-reversing a single index of the graviton propagator gives,
\begin{equation}
i\Bigl[\mbox{}_{\widehat{\mu\nu}} \Delta_{\rho\sigma}\!\Bigr] \!= 2 
\eta_{\mu (\rho} \eta_{\sigma) \nu} i\Delta_A + 4 \delta^0_{~(\mu} 
\overline{\eta}_{\nu ) (\rho} \delta^0_{~\sigma)} (i\Delta_A \!-\! i\Delta_B) 
- \frac{2 \delta^0_{~\mu} \delta^0_{~\nu} E_{\rho\sigma}}{D \!-\! 3} 
(i\Delta_A \!-\! i\Delta_C) \; . \label{tracerev1}
\end{equation}
This form is desirable because the noncovariant tensor factors multiply
differences, $(i\Delta_A - i\Delta_B)$ and $(i\Delta_A - i\Delta_C)$, which 
are only logarithmically singular at coincidence. Trace-reversing on both
indices gives,
\begin{eqnarray}
\lefteqn{ i\Bigl[\mbox{}_{\widehat{\mu\nu}} \Delta_{\widehat{\rho\sigma}}\Bigr]
= \Bigl[ 2 \eta_{\mu (\rho} \eta_{\sigma) \nu} \!-\! \eta_{\mu\nu} 
\eta_{\rho\sigma}\Bigr] i\Delta_A } \nonumber \\
& & \hspace{.7cm} + 4 \delta^0_{~(\mu} 
\overline{\eta}_{\nu ) (\rho} \delta^0_{~\sigma)} (i\Delta_A \!-\! i\Delta_B)
- 2 \Bigl( \frac{D \!-\! 2}{D \!-\! 3}\Bigr) \delta^0_{~\mu} \delta^0_{~\nu}
\delta^0_{~\rho} \delta^0_{~\sigma} (i\Delta_A \!-\! i\Delta_C) \; . \qquad 
\label{tracerev2}
\end{eqnarray}

\subsection{The Primitive 4-Point Contribution}

We might call the middle diagram of Figure~\ref{scalar} $-i M^2_{4}(x;x')$. 
From expression (\ref{MsqVEV}) we see that it involves the in-out matrix
element of,
\begin{eqnarray}
\lefteqn{ \Bigl[ \frac{i \delta^2 S[g,\phi]}{\delta \phi(x) \delta \phi(x')}
\Bigr]_{hh} = i \partial_{\mu} \Bigl[ \sqrt{-g(x)} \, g^{\mu\nu}(x) 
\partial_{\nu} \delta^D(x \!-\! x')\Bigr]_{hh} \; , } \\
& & \hspace{.8cm} = i \kappa^2 \partial_{\mu} \Bigl[ a^{D-2} \Bigl( 
\widehat{h}^{\mu \rho}(x) h^{\nu}_{~\rho}(x) \!-\! \frac14 \eta^{\mu\nu} 
\widehat{h}^{\rho\sigma}(x) h_{\rho\sigma}(x)\Bigr) \partial_{\nu} 
\delta^D(x \!-\! x') \Bigr] \; . \qquad \label{var2}
\end{eqnarray}
The matrix element involves the single trace-reversed propagator 
(\ref{tracerev1}),
\begin{equation}
-i M^2_{4} = i \kappa^2 \partial_{\mu} \Biggl\{ a^{D-2}
\Biggl( i\Bigl[\mbox{}^{\widehat{\mu\rho}} \Delta^{\nu}_{~\rho}\Bigr](x;x)
- \frac14 \eta^{\mu\nu} i \Bigl[\mbox{}^{\widehat{\rho\sigma}}
\Delta_{\rho\sigma}\Bigr](x;x) \Biggr) \partial_{\nu} \delta^D(x \!-\! x')
\Biggr\} . 
\end{equation}
It can be reduced to give,
\newpage
\begin{eqnarray}
\lefteqn{ -i M^2_{4}(x;x') = i \kappa^2 \partial_{\mu} \Biggl\{ a^{D-2} 
\Biggl( \eta^{\mu\nu} \Bigl[ -\frac{(D\!-\!5) D}{4} \, i\Delta_A(x;x) + 
\frac12 \Bigl(\frac{D \!-\!4}{D \!-\! 3}\Bigr) k\Bigr] } \nonumber \\
& & \hspace{2cm} + \delta^{\mu}_{~0} \delta^{\nu}_{~0} \Bigl[ D i\Delta_A(x;x)
+ \frac{(D \!-\!1) (D\!-\!4)}{(D\!-\!2) (D\!-\!3)} k\Bigr] \Biggr) 
\partial_{\nu} \delta^D(x \!-\! x')\Biggr\} , \qquad \label{M4final}
\end{eqnarray}
where we used the coincidence limits of $i\Delta_B$ and $i\Delta_C$ inferred 
from (\ref{DeltaBC}).

\subsection{The Primitive 3-Point Contribution}

The left hand diagram of Figure~\ref{scalar} might be called $-i M^2_{3}(x;x')$.
From expression (\ref{MsqVEV}) we see that it involves the product of two first 
variations,
\begin{equation}
\Bigl[ \frac{i \delta S[g,\phi]}{\delta \phi(x)} \Bigr]_{h \phi} = i\partial_{\mu}
\Bigl[ \sqrt{-g} \, g^{\mu\nu} \partial_{\nu} \phi\Bigr]_{h \phi} = -i \kappa
\partial_{\mu} \Bigl[ a^{D-2} \widehat{h}^{\mu\nu} \partial_{\nu} \phi\Bigr] \; .
\label{var1}
\end{equation}
The 3-point contribution involves the twice trace-reversed propagator
(\ref{tracerev2}),
\begin{equation}
-i M^2_3(x;x') = -\kappa^2 \partial_{\mu} \partial'_{\rho} \Bigl\{ (a a')^{D-2}
i\Bigl[ \mbox{}^{\widehat{\mu\nu}} \Delta^{\widehat{\rho\sigma}} \Bigr](x;x')
\, \partial_{\nu} \partial'_{\sigma} i\Delta_A(x;x') \Bigr\} . \label{M3point}
\end{equation}
Expression (\ref{tracerev2}) suggests a natural decomposition into four parts,
\begin{eqnarray}
-i M^2_{3A} &\!\!\! \equiv \!\!\!& -\kappa^2 \partial \!\cdot\! \partial'
\Bigl\{ (a a')^{D-2} i\Delta_A \!\times\! \partial \!\cdot\! \partial' 
i\Delta_A\Bigr\} \; , \qquad \label{M3A} \\
-i M^2_{3B} &\!\!\! \equiv \!\!\!& -\kappa^2 \partial^{\mu} {\partial'}^{\rho}
\Bigl\{ (a a')^{D-2} i\Delta_A \!\times\! (\partial_{\rho} \partial'_{\mu} 
\!-\! \partial_{\mu} \partial'_{\rho}) i\Delta_A \Bigr\} \; , \qquad 
\label{M3B} \\
-i M^2_{3C} &\!\!\! \equiv \!\!\!& -\kappa^2 \partial^{\mu} {\partial'}^{\rho}
\Bigl\{ (a a')^{D-2} (i\Delta_A \!-\! i\Delta_B) \nonumber \\
& & \hspace{2cm} \times \Bigl[ \overline{\eta}_{\mu\rho} \partial_0 
\partial'_0 \!-\! \delta^0_{~\mu} \overline{\partial}_{\rho} \partial'_0 \!+\! 
\delta^0_{~\rho} \overline{\partial}_{\mu} \partial_0 \!-\! \delta^0_{~\mu} 
\delta^0_{~\rho} \nabla^2 \Bigr] i\Delta_A \Bigr\} \; , \qquad \label{M3C} \\
-i M^2_{3D} &\!\!\! \equiv \!\!\!& 2\Bigl(\frac{D\!-\!2}{D\!-\!3}\Bigl) 
\kappa^2 \partial_0 \partial'_0 \Bigl\{ (a a')^{D-2} (i \Delta_A \!-\! 
i\Delta_C) \partial_0 \partial_0' i\Delta_A\Bigr\} \; . \qquad \label{M3D}
\end{eqnarray}

The terms inside the curly brackets of expression (\ref{M3A}) are 
quadratically divergent, whereas the curly brackets of (\ref{M3B}-\ref{M3D})
are only logarithmically divergent. This means we only need a few terms of
$i\Delta_A(x;x')$,
\begin{eqnarray}
\lefteqn{ i\Delta_A(x;x') = \frac{\Gamma(\frac{D}2 \!-\! 1)}{4 \pi^{\frac{D}2}
(a a')^{\frac{D}2 -1}} \Biggl\{ \frac1{\Delta x^{D-2}} + \frac{D (D\!-\!2)}{8 
(D\!-\!4)} \frac{a a' H^2}{\Delta x^{D-4}} + \dots\Biggr\} } \nonumber \\
& & \hspace{5.5cm} + k \Biggl\{ -\pi {\rm cot}\Bigl( \frac{D \pi}{2}\Bigr) +
\ln(a a') + \dots \Biggr\} . \qquad \label{DAexpand}
\end{eqnarray}
The 4-step procedure for reducing expressions (\ref{M3A}-\ref{M3D}) is,
\begin{enumerate}
\item{Act the two inner derivatives on $i\Delta_A$;\footnote{Note that this 
can produce a delta function when acting on the most singular term,
\begin{eqnarray}
\partial_{\alpha} \partial'_{\beta} \Bigl[\frac1{\Delta x^{D-2}}\Bigr] 
= \frac{\delta^0_{~\alpha} \delta^0_{~\beta} 4 \pi^{\frac{D}2} i 
\delta^D(x \!-\! x')}{\Gamma(\frac{D}2 \!-\! 1)} + (D \!-\! 2) \Bigl[
\frac{\eta_{\alpha\beta}}{\Delta x^D} - \frac{D \Delta x_{\alpha} 
\Delta x_{\beta}}{\Delta x^{D+2}} \Bigr] \; . \nonumber 
\end{eqnarray}}}
\item{Multiply by the scalar propagators from the graviton propagator,
and retain only those terms which are nonzero in the unregulated limit;}
\item{Extract derivatives from the quadratically divergent terms to reach
a logarithmically divergent form,
\begin{equation}
\frac1{\Delta x^{2D-2}} = \frac{\partial^2}{2 (D\!-\!2)^2} \Bigl[ 
\frac1{\Delta x^{2D-4}} \Bigr] \qquad {\rm ; and}
\end{equation}}
\item{Localize the divergence and take the unregulated limit on the 
remainder using the flat space propagator equation \cite{Onemli:2002hr},
\begin{eqnarray}
\lefteqn{\frac1{\Delta x^{2D-4}} = \frac{\partial^2}{2 (D\!-\!3) (D\!-\!4)}
\Bigl[ \frac1{\Delta x^{2D-6}}\Bigr] \; , } \\
& & \hspace{-0.7cm} = \frac{\mu^{D-4} 4 \pi^{\frac{D}2} i\delta^D(x \!-\! x')}{
2 (D\!-\!3) (D\!-\!4) \Gamma(\frac{D}2 \!-\!1)} + \frac{\partial^2}{2 (D\!-\!3) 
(D\!-\!4)} \Bigl[ \frac1{\Delta x^{2D-6}} \!-\! \frac{\mu^{D-4}}{\Delta x^{D-2}}
\Bigr] , \qquad \\
& & \hspace{-0.7cm} = \frac{\mu^{D-4} 4 \pi^{\frac{D}2} i\delta^D(x \!-\! x')}{
2 (D\!-\!3) (D\!-\!4) \Gamma(\frac{D}2 \!-\!1)} - \frac14 \partial^2 \Bigl[ 
\frac{\ln(\mu^2 \Delta x^2)}{\Delta x^2} \Bigr] + O(D \!-\! 4) \; . 
\label{locdiv}
\end{eqnarray}}
\end{enumerate}

Before applying the 4-step procedure on $-i M^2_{3B}(x;x')$ it is useful 
to expand the traces over $\mu$ and $\rho$,
\begin{equation}
-i M^2_{3B}(x;x') = -\kappa^2 (\partial_0 \!+\! \partial'_0) \partial_i
\Bigl\{ (a a')^{D-2} i\Delta_A(x;x') (\partial_0 \!+\! \partial'_0) \partial_i
i \Delta_A(x;x') \Bigr\} \; . \label{M2Bprime}
\end{equation}
A similar expansion of $-i M^2_{3C}(x;x')$ gives,
\begin{eqnarray}
\lefteqn{-i M^2_C = \kappa^2 \nabla^2 \Bigl\{ (a a')^{D-2} (i\Delta_A \!-\!
i\Delta_B) \partial_0 \partial'_0 i\Delta_A \Bigr\} + \kappa^2 \partial_0 
\partial_i \Bigl\{ (a a')^{D-2} } \nonumber \\
& & \hspace{0cm} \times (i\Delta_A \!-\! i\Delta_B) \partial'_0 \partial_i 
i\Delta_A \Bigr\} + \kappa^2 \partial'_0 \partial_i \Bigl\{ (a a')^{D-2} 
(i \Delta_A \!-\! i \Delta_B) \partial_0 \partial_i i\Delta_A\Bigr\} \qquad 
\nonumber \\
& & \hspace{5cm} + \kappa^2 \partial_0 \partial'_0 \Bigl\{ (a a')^{D-2}
(i\Delta_A \!-\! i\Delta_B) \nabla^2 i\Delta_A \Bigr\} . \qquad
\label{M2Cprime}
\end{eqnarray}
Also note that each divergence is proportional to one of two constants,
\newpage
\begin{equation}
A_0 \equiv \kappa^2 k \pi {\rm cot}\Bigl(\frac{D\pi}{2}\Bigr) \qquad , \qquad
A_1 \equiv \frac{\kappa^2 H^2}{4 \pi^{\frac{D}2}} \frac{\mu^{D-4} 
\Gamma(\frac{D}2)}{(D \!-\!3) (D\!-\!4)} \; . \label{A0A1}
\end{equation}
Now apply the 4-step procedure to find,
\begin{eqnarray}
\lefteqn{ -i M^2_{3A}(x;x') = \Bigl[A_0 + \frac{_3}{^4} A_1\Bigr] \partial^{\mu} 
\Bigl[ a^2 \partial_{\mu} i \delta^D(x \!-\! x')\Bigr] + \frac{\kappa^2 H^2 
\partial \!\cdot\! \partial'}{64 \pi^4} \Bigl\{ a a' \partial_0^2 \Bigl[ 
\frac1{\Delta x^2}\Bigr]} \nonumber \\
& & \hspace{0cm} + \frac{_3}{^2} a a' \partial^2 \Bigl[ \frac{\ln(\mu^2 \Delta x^2)
}{\Delta x^2}\Bigr] + 4 a^2 {a'}^2 H^2 \Bigl[ \frac{\ln(\frac14 H^2 \Delta x^2) 
\!+\! 1}{\Delta x^2}\Bigr] \nonumber \\
& & \hspace{3.5cm} + \frac{_1}{^2} a^2 {a'}^2 H^2 \partial_0^2 \Bigl[
\ln^2( \frac{_1}{^4} H^2 \Delta x^2) \!+\! 3 \ln( \frac{_1}{^4} H^2 \Delta x^2 )
\Bigr] \Bigr\} , \label{M3Atotal} \qquad \\
\lefteqn{ -i M^2_{3B}(x;x') = \frac{_1}{^2} A_1 \partial_i \Bigl[ a^2 \partial_i 
i\delta^D(x \!-\! x')\Bigr] - \frac{\kappa^2 H^2 \nabla^2}{64 \pi^4} \Bigl\{
a a' \partial^2 \Bigl[ \frac{\ln(\mu^2 \Delta x^2)}{\Delta x^2}\Bigr] } 
\nonumber \\
& & \hspace{0cm} + 12 a^2 {a'}^2 H^2 \Bigl[ \frac{\ln(\frac14 H^2 \Delta x^2) 
\!+\! \frac32}{\Delta x^2}\Bigr] + \frac{_1}{^2} a^2 {a'}^2 H^2 \Delta \eta^2 
\partial^2 \Bigl[ \frac{\ln(\mu^2 \Delta x^2)}{\Delta x^2}\Bigr] \nonumber \\
& & \hspace{5.5cm} + 4 a^3 {a'}^3 H^4 \Delta \eta^2 \Bigl[ \frac{\ln(\frac14 H^2 
\Delta x^2) \!+\! \frac32}{\Delta x^2}\Bigr] \Bigr\} , \label{M3Btotal} 
\qquad \\
\lefteqn{ -i M^2_{3C}(x;x') = A_1 \Bigl\{ -\frac{_1}{^4} \partial_i \Bigl[ a^2 
\partial_i i\delta^D(x \!-\!x') \Bigr] + \frac{_1}{^4} ( D \!-\! 1) \partial_0 
\Bigl[ a^2 \partial_0 i \delta^D(x \!-\!x') \Bigr] \Bigr\} } \nonumber \\
& & \hspace{-0.7cm} + \frac{\kappa^2 H^2 \nabla^2}{64 \pi^4} \Bigl\{ \frac{_1}{^2}
a a' \partial^2 \Bigl[ \frac{\ln(\mu^2 \Delta x^2)}{\Delta x^2}\Bigr] + 6 a a' 
\partial_0^2 \Bigl[ \frac{\ln(\frac14 H^2 \Delta x^2) \!+\! 2}{\Delta x^2}\Bigr] 
\nonumber \\
& & \hspace{0cm} + \frac{_1}{^2} a^2 {a'}^2 H^2 \partial^2 \Bigl[ \ln^2(
\frac{_1}{^4} H^2 \Delta x^2) \!+\! \frac{_1}{^2} \ln(\frac{_1}{^4} H^2 
\Delta x^2)\Bigr] \!-\! a^2 {a'}^2 H^2 \partial_0^2 \Bigl[\ln^2( \frac{_1}{^4} 
H^2 \Delta x^2) \nonumber \\
& & \hspace{0.5cm} + \frac{_7}{^2} \ln(\frac{_1}{^4} H^2 \Delta x^2) \Bigr] + 
a^3 {a'}^3 H^4 \Bigl[ \ln^2( \frac{_1}{^4} H^2 \Delta x^2) \!+\! 3 
\ln(\frac{_1}{^4} H^2 \Delta x^2) \Bigr] \nonumber \\
& & \hspace{2.5cm} + \frac{_1}{^2} a^3 {a'}^3 H^4 \partial_0^2 \Bigl( \Delta x^2 
\Bigl[ \ln^2( \frac{_1}{^4} H^2 \Delta x^2 ) \!+\! \ln( \frac{_1}{^4} H^2 
\Delta x^2) \!-\! 1\Bigr] \Bigr) \Bigr\} \nonumber \\
& & \hspace{-0.7cm} + \frac{\kappa^2 H^2 \partial_0 \partial_0'}{64 \pi^4} \Bigl\{
\frac{_3}{^2} a a' \partial^2 \Bigl[ \frac{\ln(\mu^2 \Delta x^2)}{\Delta x^2}
\Bigr] \!-\! 2 a a' \nabla^2 \Bigl[ \frac{\ln(\frac14 H^2 \Delta x^2) 
\!+\! 2}{\Delta x^2} \Bigr] \!-\! \frac{_3}{^2} a^2 {a'}^2 H^2 \nonumber \\
& & \hspace{0.3cm} \times \partial^2 \ln( \frac{_1}{^4} H^2 \Delta x^2) \!+\! 
\frac{_1}{^2} a^2 {a'}^2 H^2 \nabla^2 \Bigl[ \ln^2(\frac{_1}{^4} H^2 \Delta x^2) 
\!+\! 3 \ln( \frac{_1}{^4} H^2 \Delta x^2) \Bigr] \Bigr\} , \qquad  
\label{M3Ctotal} \\
\lefteqn{ -i M^2_{3D}(x;x') = A_1 \partial_0 \Bigl[ a^2 \partial_0 i 
\delta^D(x \!-\!x') \Bigr] + \frac{\kappa^2 H^2 \partial_0 \partial_0'}{64 \pi^4} 
\Bigl\{2 a a' \partial^2 \Bigl[ \frac{\ln(\mu^2 \Delta x^2)}{\Delta x^2}\Bigr] }
\nonumber \\
& & \hspace{7cm} + 8 a a' \partial_0^2 \Bigl[ \frac{\ln(\frac14 H^2 \Delta x^2) 
\!+\! 2}{\Delta x^2} \Bigr] \Bigr\} . \qquad \label{M3Dtotal}
\end{eqnarray}

\subsection{Renormalization}

Reducing the 4-point contribution (\ref{M4final}) to the 3-point form 
(\ref{M3Atotal}-\ref{M3Dtotal}) gives,
\newpage
\begin{equation}
-i M^2_{4}(x;x') = \frac{_1}{^4} D(D \!-\! 5) A_0 \partial^{\mu} \Bigl[a^2
\partial_{\mu} i \delta^D(x \!-\! x')\Bigr] - D A_0 \partial_0 \Bigl[ a^2
\partial_0 i \delta^D(x \!-\! x') \Bigr] \; . \label{M4total}
\end{equation}
We can now sum the divergent parts from expressions 
(\ref{M3Atotal}-\ref{M4total}),
\begin{eqnarray}
\lefteqn{-i M^2_{\rm div}(x;x') = \Bigl[ \frac{_1}{^4} (D\!-\!1) (D\!-\!4) A_0 \!+\!
A_1\Bigr] \partial^{\mu} \Bigl[a^2 \partial_{\mu} i \delta^D(x \!-\! x')\Bigr] }
\nonumber \\
& & \hspace{4cm} + \Bigl[-D A_0 \!+\! \frac{_1}{^4} (D \!+\! 4) A_1 \Bigr]
\partial_0 \Bigl[ a^2 \partial_0 i \delta^D(x \!-\! x') \Bigr] \; . \qquad 
\label{Mdiv}
\end{eqnarray}
Recall that $A_0$ and $A_1$ were defined in (\ref{A0A1}).

The 1-loop divergences (\ref{Mdiv}) are canceled by three counterterms, 
\begin{equation}
\Delta \mathcal{L} = -\frac12 \alpha_1 \square \phi \square \phi \sqrt{-g}
- \frac12 \alpha_2 R \partial_{\mu} \phi \partial_{\nu} \phi g^{\mu\nu} 
\sqrt{-g} - \frac12 \alpha_3 R \partial_0 \phi \partial_0 \phi g^{00} 
\sqrt{-g} \; . \label{DeltaL}
\end{equation}
Hence the final diagram of Figure~\ref{scalar} is, 
\begin{eqnarray}
\lefteqn{-i M^2_{\rm ctm}(x;x') \equiv \Bigl[\frac{i \delta^2 \Delta S}{
\delta \phi(x) \delta \phi(x')} \Bigr]_{1} = -\alpha_1 \partial^{\mu} 
{\partial'}^{\rho} \Biggl[ (a a')^{D-2} \partial_{\mu} \partial'_{\rho} 
\Bigl( \frac{i \delta^D(x \!-\! x')}{a^D} \Bigr) \Biggr] } \nonumber \\
& & \hspace{1.5cm} + \alpha_2 \partial^{\mu} \Bigl[ R a^{D-2} \partial_{\mu} 
i\delta^D(x \!-\!x') \Bigr] - \alpha_3 \partial_0 \Bigl[ R a^{D-2} 
\partial_0 i\delta^D(x \!-\! x') \Bigr] \; , \qquad \label{cterms}
\end{eqnarray} 
where the Ricci scalar is $R = D (D-1) H^2$. Comparison between expressions
(\ref{Mdiv}) and (\ref{cterms}) implies,
\begin{eqnarray}
\alpha_1 &\!\!\! = \!\!\!& 0 \; , \label{alpha1} \\
\alpha_2 R &\!\!\! = \!\!\!& -\frac{\kappa^2 H^2 \mu^{D-4}}{4 \pi^{\frac{D}2}}
\Bigl\{ \frac{\Gamma(D)}{\Gamma(\frac{D}2)} \frac{(D \!-\!4)}{16}
\pi {\rm cot}\Bigl( \frac{D\pi}{2}\Bigr) + \frac{\Gamma(\frac{D}2)}{(D\!-\!3)
(D\!-\!4)} \Bigr\} , \qquad \label{alpha2} \\
\alpha_3 R &\!\!\! = \!\!\!& -\frac{\kappa^2 H^2 \mu^{D-4}}{4 \pi^{\frac{D}2}}
\Bigl\{ \frac{\Gamma(D\!-\!1)}{\Gamma(\frac{D}2)} \frac{D}{4} \pi 
{\rm cot}\Bigl(\frac{D \pi}{2}\Bigr) - \frac{(D \!+\! 4) \Gamma(\frac{D}2)}{
4 (D\!-\!3) (D\!-\!4)} \Bigr\} . \qquad \label{alpha3}
\end{eqnarray}
The vanishing of $\alpha_1$ is an artifact of $\alpha = \beta = 1$ gauge in
the flat space limit (\ref{M0}). The counterterms proportional to $\alpha_2$ 
and $\alpha_3$ vanish in the flat space limit and their potential gauge 
dependence is not known.

Combining the primitive divergence with the counterterms and taking the
unregulated limit gives,
\begin{eqnarray}
\lefteqn{-i M^2_{\rm div}(x;x') -i M^2_{\rm ctm}(x;x') = -
\frac{\kappa^2 H^2}{4 \pi^2} \partial^{\mu} \Bigl[ a^2 \ln(a) \partial_{\mu}
i \delta^4(x \!-\! x')\Bigr] } \nonumber \\
& & \hspace{3cm} + \frac{\kappa^2 H^2}{2 \pi^2} \partial_0 \Bigl[ a^2 
\ln\Bigl( \frac{4 \mu^2 a}{H^2}\Bigr) \partial_0 i \delta^4(x \!-\! x')\Bigr]
+ O(D \!-\! 4) \; . \qquad \label{localterms}
\end{eqnarray}
The renormalized self-mass comes from adding these local terms to the
nonlocal parts of (\ref{M3Atotal}-\ref{M3Dtotal}), and then simplifying
the sum,
\begin{eqnarray}
\lefteqn{ -i M^2_{\rm ren}(x;x') = -\frac{\kappa^2 H^2}{4 \pi^2} 
\partial^{\mu} \Bigl[ a^2 \ln(a) \partial_{\mu} i \delta^4(x \!-\! x')
\Bigr] } \nonumber \\
& & \hspace{0cm} + \frac{\kappa^2 H^2}{2 \pi^2} \partial_0 \Bigl[ a^2 
\ln\Bigl( \frac{4 \mu^2 a}{H^2}\Bigr) \partial_0 i \delta^4(x \!-\! x')
\Bigr] + \frac{\kappa^2 H^2 \partial_0 \partial_0'}{64 \pi^4} \Biggr\{ 
a a' \partial_0 \partial_0' \Bigl[ \frac1{\Delta x^2} \Bigr] 
\nonumber \\
& & \hspace{0cm} + 2 a a' (\partial_0 \partial_0' \!+\! \nabla^2) 
\Bigl[ \frac{\ln(\mu^2 \Delta x^2)}{\Delta x^2}\Bigr] - 2 a a' 
(4 \partial_0 \partial_0' \!+\! \nabla^2) \Bigl[ 
\frac{\ln(\frac14 H^2 \Delta x^2) \!+\! 2}{\Delta x^2} \Bigr] \Biggr\} 
\nonumber \\
& & \hspace{-0.7cm} + \frac{\kappa^2 H^2 \nabla^2}{64 \pi^4} \Biggl\{ -2 a a'
(\partial_0 \partial_0' \!+\! \nabla^2) \Bigl[ \frac{\ln(\mu^2 \Delta x^2)}{
\Delta x^2}\Bigr] - 6 a a' \partial_0 \partial_0' \Bigl[ \frac{\ln(\frac14 H^2 
\Delta x^2) \!+\! \frac{11}{6}}{\Delta x^2} \Bigr] \nonumber \\
& & \hspace{0cm} + \frac{_1}{^4} a^2 {a'}^2 H^2 (\partial_0 \partial_0' \!-\! 
\nabla^2) \ln(\mu^2 \Delta x^2) + \frac{_{15}}{^4} a^2 {a'}^2 H^2 \partial_0 
\partial_0' \ln(\frac{_1}{^4} H^2 \Delta x^2) \nonumber \\
& & \hspace{3cm} - a^2 {a'}^2 H^2 \nabla^2 \Bigl[ \frac{_3}{^2} 
\ln^2(\frac{_1}{^4} H^2 \Delta x^2) + \frac{_5}{^4} \ln( \frac{_1}{^4} H^2 
\Delta x^2)\Bigr] \Biggr\} . \qquad \label{Mren}
\end{eqnarray}

\section{The Linearized Effective Field Equation}

The purpose of this section is to use the renormalized self-mass (\ref{Mren})
to quantum-correct the linearized effective field equation and then solve
this equation for scalar radiation and for the scalar exchange potential.
We begin by explaining the Schwinger-Keldysh formalism that is used to 
produce a causal and real effective field equation. The equation is then
solved perturbatively, first for scalar radiation and then for the exchange 
potential. The section closes by using the renormalization group to explain
the latter.  

\subsection{Schwinger-Keldysh Formalism}

The linearized effective field equation is,
\begin{equation}
\sqrt{-g} \square \phi(x) = \partial^{\mu} \Bigl[ a^2 \partial_{\mu} \phi(x)
\Bigr] \equiv \mathcal{D} \phi(x) = J(x) + \int \!\! d^4x' M^2(x;x') \phi(x') 
\; , \label{phieqn}
\end{equation}
where $J(x)$ is the source. Substituting expression (\ref{Mren}) for the
self-mass results in an equation with three peculiar properties:
\begin{itemize}
\item{It isn't local because $M^2_{\rm ren}(x;x')$ fails to vanish for 
${x'}^{\mu} \neq x^{\mu}$;}
\item{It isn't causal because $M^2_{\rm ren}(x;x')$ fails to vanish for
${x'}^{\mu}$ outside the past light-cone of $x^{\mu}$; and}
\item{It isn't real because $M^2_{\rm ren}(x;x')$ has a nonzero imaginary
part.}
\end{itemize}
Effective field equations are unavoidably nonlocal but the other two 
properties derive from $-i M^2_{\rm ren}(x;x')$ representing an in-out
amplitude rather than a true expectation value. Of course that is what
the Feynman rules produce, and it is exactly the right thing for 
scattering amplitudes. However, the ``in'' and ``out'' vacua disagree due 
to the very cosmological particle production (\ref{occunum}) whose effect 
we seek to study, and causality precludes the $S$-matrix from being an 
observable. It is therefore more sensible to study the evolution of the 
expectation value of $\phi(x)$ in the presence of a state which was 
empty in the distant past. The Schwinger-Keldysh formalism provides a
diagrammatic procedure for computing this which is almost as simple to 
use as the Feynman rules \cite{Schwinger:1960qe,Mahanthappa:1962ex,
Bakshi:1962dv,Bakshi:1963bn,Keldysh:1964ud}. This expectation value 
obeys the Schwinger-Keldysh effective field equations, which are both 
causal and real \cite{Chou:1984es,Jordan:1986ug,Calzetta:1986ey}.

It is straightforward to convert the in-out effective field equations
to the in-in equations of the Schwinger-Keldysh formalism. The rules
are \cite{Ford:2004wc}:
\begin{itemize}
\item{End points of lines in the diagrammatic formalism have $\pm$ 
polarizations, resulting in four propagators and $2^N$ 1PI $N$-point 
functions;}
\item{The $++$ propagator is the same as the Feynman propagator, and
the $--$ propagator is its complex conjugate;}
\item{The $+-$ and $-+$ propagators are homogeneous solutions of the 
propagator equation, which are obtained from the Feynman propagator by 
changing the $i\epsilon$ in the conformal coordinate interval from 
(\ref{ydef}) to,
\begin{eqnarray}
\Delta x^2_{+-} & \equiv & \Vert \vec{x} \!-\! \vec{x}' \Vert^2 - 
(\eta \!-\! \eta' \!+\! i \epsilon)^2 \; , \label{x+-def} \\
\Delta x^2_{-+} & \equiv & \Vert \vec{x} \!-\! \vec{x}' \Vert^2 - 
(\eta \!-\! \eta' \!-\! i \epsilon)^2 \qquad {\rm ; and} \label{x-+def}
\end{eqnarray}}
\item{Vertices carry only a single polarity, so all their lines are either 
$+$ or $-$, with the $+$ vertices being the same as those of the Feynman 
rules and the $-$ vertices being their complex conjugates.}
\end{itemize}

The term ``$M^2(x;x')$'' in the Schwinger-Keldysh effective field 
equation (\ref{phieqn}) is $M^2_{++}(x;x') + M^2_{+-}(x;x')$. It is
real because the $\pm$ vertex at ${x'}^{\mu}$ results in a relative
minus sign, and because $\Delta x^2_{+-} = \Delta x^2_{++}$ for $\eta 
< \eta'$ whereas $\Delta x^2_{+-} = (\Delta x^2_{++})^*$ for $\eta > 
\eta'$. To see causality one first eliminates inverse powers of 
$\Delta x^2$,
\begin{eqnarray}
\frac1{\Delta x^2} & = & \frac{_1}{^4} \partial^2 \ln(\mu^2 \Delta x^2) 
\; , \\
\frac{\ln(\mu^2 \Delta x^2)}{\Delta x^2} & = & \frac{_1}{^8} \partial^2
\Bigl[ \ln^2(\mu^2 \Delta x^2) - 2 \ln(\mu^2 \Delta x^2)\Bigr] \; .
\end{eqnarray}
Now note that differences of powers of $++$ and $+-$ logarithms are
proportional to $\theta(\Delta \eta - \Delta r)$, where $\Delta \eta 
\equiv \eta - \eta'$ and $\Delta r \equiv \Vert \vec{x} - \vec{x}'\Vert$,
\begin{eqnarray}
\ln(\mu^2 \Delta x^2_{++}) - \ln(\mu^2 \Delta x^2_{+-}) & = & 2\pi i 
\theta(\Delta \eta \!-\! \Delta r) \; , \\
\ln^2(\mu^2 \Delta x^2_{++}) - \ln^2(\mu^2 \Delta x^2_{+-}) & = & 4\pi i 
\theta(\Delta \eta \!-\! \Delta r) \ln[\mu^2 (\Delta \eta^2 \!-\! 
\Delta r^2)] \; .
\end{eqnarray}
In converting expression (\ref{Mren}) to Schwinger-Keldysh form we will
employ the notation $\Theta \equiv \theta(\Delta \eta - \Delta r)$ 
to achieve a more compact form,
\begin{eqnarray}
\lefteqn{ M^2_{\rm SK}(x;x') = \frac{\kappa^2 H^2}{4 \pi^2} \Biggl\{
\partial^{\mu} \Bigl[a^2 \ln(a) \partial_{\mu}\Bigr] \!-\! 2 \partial_0
\Bigl[ a^2 \ln\Bigl( \frac{4 \mu^2 a}{H^2}\Bigr) \partial_0 \Bigr] 
\Biggr\} \delta^4(x \!-\! x') } \nonumber \\
& & \hspace{-0.5cm} + \frac{\kappa^2 H^2 \partial_0 \partial'_0}{128 \pi^3}
\Biggl\{-2 a a' (\partial_0 \partial'_0 \!+\! \nabla^2) \partial^2 \Bigl[ 
\Theta \ln[\mu^2 (\Delta \eta^2 \!-\! \Delta r^2)] \Bigr] + 9 a a' 
\partial_0 \partial'_0 \partial^2 \Theta \nonumber \\
& & \hspace{2cm} + 4 a a' \nabla^2 \partial^2 \Theta + 2 a a' 
(4 \partial_0 \partial'_0 \!+\! \nabla^2) \partial^2 \Bigl[ \Theta 
\ln[\frac{_1}{^4} H^2 (\Delta \eta^2 \!-\! \Delta r^2)] \Bigr] \Biggr\} 
\nonumber \\
& & \hspace{-0.5cm} + \frac{\kappa^2 H^2 \nabla^2}{128 \pi^3} \Biggl\{
2 a a' (\partial_0 \partial'_0 \!+\! \nabla^2) \partial^2 \Bigl[ \Theta
\ln[\mu^2 (\Delta \eta^2 \!-\! \Delta r^2)] \Bigr] + 3 a a' \partial_0
\partial'_0 \partial^2 \Theta \nonumber \\
& & \hspace{0cm} - 2 a a' \nabla^2 \partial^2 \Theta + 6 a a' \partial_0 
\partial'_0 \partial^2 \Bigl[\Theta \ln[\frac{_1}{^4} H^2 (\Delta \eta^2 
\!-\! \Delta r^2)] \Bigr] - 16 a^2 {a'}^2 H^2 \partial_0 \partial'_0 
\Theta \nonumber \\
& & \hspace{1.8cm} + 6 a^2 {a'}^2 H^2 \nabla^2 \Theta + 12 a^2 {a'}^2 H^2 
\nabla^2 \Bigl[ \Theta \ln[\frac{_1}{^4} H^2 (\Delta \eta^2 \!-\! 
\Delta r^2)] \Bigr] \Biggr\} . \qquad \label{MSK}
\end{eqnarray}

\subsection{The Scalar Mode Function}

Scalar radiation corresponds to $J(x) = 0$ and solutions take the form,
\begin{equation}
\phi(x) = u(\eta,k) e^{i \vec{k} \cdot \vec{x}} \qquad , \qquad 
k \equiv \Vert \vec{k} \Vert \; . \label{radform}
\end{equation}
The spatial exponential can be factored out using translation invariance,
\begin{equation}
\mathcal{D} u(\eta,k) \equiv -a^2 \Bigl[\partial_0^2 + 2 a H \partial_0
+ k^2\Bigr] u(\eta,k) = \int \!\! d^4x' M^2_{\rm SK}(x;x') u(\eta',k)
e^{-i \vec{k} \cdot \Delta \vec{x}} . \label{modeeqn}
\end{equation}
Here any spatial derivatives in the self-mass are replaced by 
$\partial_i \longrightarrow i k_i$.

Because we only have the 1-loop self-mass, equation (\ref{modeeqn}) must 
be solved perturbatively ($u = u_0 + u_1 + \dots$) in powers of the 
loop-counting parameter $\kappa^2 \equiv 16 \pi G$. The tree order solution 
is,
\begin{equation}
u_0(\eta,k) = \frac{H}{\sqrt{2 k^3}} \Bigl[1 - \frac{i k}{a H}\Bigr]
\exp\Bigl[ \frac{i k}{a H} \Bigr] \xrightarrow{a \to \infty} 
\frac{H}{\sqrt{2 k^3}} \Bigl[1 + \frac{k^2}{2 a^2 H^2} + \dots \Bigr] \; , 
\label{treemode}
\end{equation}
and it is useful to note,
\begin{equation}
\partial_0 u_0(\eta,k) = \frac{H}{\sqrt{2 k^3}} \Bigl[-\frac{k^2}{a H} \Bigr]
\exp\Bigl[ \frac{i k}{a H}\Bigr] \quad \Longrightarrow \quad (\partial_0^2 
+ k^2) \Bigl[ a \partial_0 u_0(\eta,k)\Bigr] = 0 \; . \label{keyderiv}
\end{equation}
Relation (\ref{keyderiv}) means that the 2nd and 3rd lines of expression
(\ref{MSK}) for $M^2_{\rm SK}$ make no contribution to the 1-loop  
correction,
\begin{eqnarray}
\lefteqn{\mathcal{D} u_1(\eta,k) = \frac{\kappa^2 H^2}{4 \pi^2} \Bigl\{
-3 a^3 H \partial_0 u_0(\eta,k) + 2 \ln\Bigl( \frac{4 \mu^2 a}{H^2}\Bigr) 
a^2 k^2 u_0(\eta,k) \Bigr\} } \nonumber \\
& & \hspace{-0.5cm} - \frac{\kappa^2 H^4 k^2 a}{64 \pi^3} \Biggl\{2
(\partial_0^2 \!+\! k^2) \! \int \!\! d^4x' \Theta \ln[\mu^2 (\Delta \eta^2
\!-\! \Delta r^2)] {a'}^3 u_0(\eta',k) e^{-i \vec{k} \cdot \Delta \vec{x}} 
\nonumber \\
& & \hspace{0cm} + (3 \partial_0^2 \!-\! 2 k^2) \!\! \int \!\! d^4x' 
\Theta {a'}^3 u_0(\eta',k) e^{-i \vec{k} \cdot \Delta \vec{x}} \!+\! 6 
\partial_0^2 \!\! \int \!\! d^4x' \Theta \ln[\frac{_1}{^4} H^2 
(\Delta \eta^2 \!-\! \Delta r^2)] \nonumber \\
& & \hspace{1.5cm} \times {a'}^3 u_0(\eta',k) e^{-i \vec{k} \cdot 
\Delta \vec{x}} + a (8 \partial_0^2 \!-\! 3 k^2) \!\! \int \!\! d^4x' 
\Theta {a'}^2 u_0(\eta',k) e^{-i \vec{k} \cdot \Delta \vec{x}} \nonumber \\
& & \hspace{2.5cm} - 6 a k^2 \!\! \int \!\! d^4x' \Theta \ln[\frac{_1}{^4} 
H^2 (\Delta \eta^2 \!-\! \Delta r^2)] {a'}^2 u_0(\eta',k) e^{-i \vec{k} 
\cdot \Delta \vec{x}} \Biggr\} . \qquad \label{u1eqn}
\end{eqnarray}
In reaching this form we have also used,
\begin{equation}
(\partial_0^2 \!+\! k^2) \Bigl[a u_0(\eta,k)\Bigr] = 2 a^3 H^2 u_0(\eta,k)
\; . \label{2ndkey}
\end{equation}

Equation (\ref{u1eqn}) has the general form,
\begin{equation}
- a^2 \Bigl[ \partial_0^2 + 2 a H \partial_0 + k^2\Bigr] u_1(\eta,k) = S(\eta) 
\; . \label{sourcedef}
\end{equation}
It is important to understand the relation between the asymptotic late time 
form of the source $S(\eta)$ and the late time form it induces in 
$u_1(\eta,k)$,
\begin{eqnarray}
S = a^4 H^2 \ln(a) \Longrightarrow u_1 \rightarrow -\frac{_1}{^6} \ln^2(a) 
& , & S = a^4 H^2 \Longrightarrow u_1 \rightarrow -\frac{_1}{^3} \ln(a) \; , 
\qquad \label{S4} \\
S = a^3 H^2 \ln(a) \Longrightarrow u_1 \rightarrow +\frac{\ln(a)}{2 a} 
& , & S = a^3 H^2 \Longrightarrow u_1 \rightarrow +\frac{1}{2 a} \; , \qquad 
\label{S3} \\
S = a^2 H^2 \ln(a) \Longrightarrow u_1 \rightarrow +\frac{\ln(a)}{2 a^2} 
& , & S = a^2 H^2 \Longrightarrow u_1 \rightarrow +\frac{1}{2 a^2} \; . 
\qquad \label{S2}
\end{eqnarray} 
One can see that the leading asymptotic form of the local source terms on the
first line of (\ref{u1eqn}) is,
\begin{equation}
\frac{\kappa^2 H^2}{4 \pi^2} \Bigl\{-3a^3 H \partial_0 u_0 + 2 
\ln\Bigl( \frac{4 \mu^2 a}{H^2}\Bigr) a^2 k^2 u_0 \Bigr\}
\longrightarrow \frac{\kappa^2 H^2}{4 \pi^2} \times 2 a^2 \ln(a) k^2 u_0(0,k)
\; . \label{latelocal}
\end{equation}
Because all the nonlocal terms contain at least one factor of $k^2$, we can
suppress higher powers of $k^2$, which carry factors of $1/a^2$. This simplifies 
the integrations, 
\begin{eqnarray}
\lefteqn{ -\frac{\kappa^2 H^4 k^2 u_0(0,k) a}{64 \pi^3} \Biggl\{ 2 \partial_0^2 
\!\! \int \!\! d^4x' \Theta \ln[ \mu^2 (\Delta \eta^2 \!-\! \Delta r^2)] {a'}^3
+ 3 \partial_0^2 \!\! \int \!\! d^4x' \Theta {a'}^3 } \nonumber \\
& & \hspace{3cm} + 6 \partial_0^2 \!\! \int \!\! d^4x' \Theta \ln[ \frac{_1}{^4}
H^2 (\Delta \eta^2 \!-\! \Delta r^2)] {a'}^3 + 8 a \partial_0^2 \!\! \int \!\!
d^4x' \Theta {a'}^2 \Biggr\} \nonumber \\
& & \hspace{0cm} \longrightarrow \frac{\kappa^2 H^2}{4 \pi^2} \times 
\Bigl[\ln\Bigl(\frac{H}{2 \mu}\Bigr)\Bigr] \!+\! \frac54\Bigr] a^2 k^2 u_0(0,k) 
\; . \qquad \label{latenonlocal} 
\end{eqnarray}
In view of (\ref{S2}), expressions (\ref{latelocal}) and (\ref{latenonlocal})
imply,
\begin{equation}
u_1(\eta,k) \longrightarrow \frac{\kappa^2 H^2}{4 \pi^2} \!\times\! 
\frac{k^2 \ln(a)}{a^2 H^2} \!\times\! u_0(0,k) \; . \label{u1final}
\end{equation}
Hence 1-loop graviton corrections do not change the constant freeze-in value of 
the mode function (\ref{treemode}), but they do slow down the rate of approach 
to this constant.

\subsection{The Response to a Point Source}

The scalar exchange potential corresponds to $J(x) = K a \delta^3(\vec{x})$.
The solution has the form $\Phi(\eta,r)$, so this system is fundamentally 
2-dimensional, unlike the 1-dimensional problem of the mode function $u(\eta,k)$.
The tree order response is \cite{Glavan:2019yfc},
\begin{equation}
\Phi_0(t,r) = \frac{K H}{4 \pi} \Bigl\{ -\frac1{a H r} + 
\ln\Bigl(H r \!+\! \frac1{a}\Bigr) \Bigr\} \xrightarrow{a \to \infty}
\frac{H K}{4 \pi} \Bigl\{ \ln(H r) - \frac1{2 a^2 H^2 r^2} + \dots\Bigr\}
\; . \label{treepot}
\end{equation} 
Derivatives of $\Phi_0(\eta,r)$ are,
\begin{equation}
\partial_0 \Phi_0(\eta,r) = \frac{K H^2}{4\pi} \Bigl\{ \frac1{H r} - 
\frac1{Hr \!+\! \frac1{a}} \Bigr\} \quad , \quad \partial_0^2 \Phi_0(\eta,r)
= -\frac{K H^3}{4 \pi} \frac1{(H r \!+\! \frac1{a})^2} \; , \label{tderiv}
\end{equation}
\begin{equation}
\nabla^2 \Phi_0(\eta,r) =  \frac{K \delta^3(\vec{x})}{a} + \frac1{a^2} 
\partial_0 \Bigl[ a^2 \partial_0 \Phi_0\Bigr] . \label{rderiv}
\end{equation}
In addition to $\mathcal{D} \Phi_0 = K a \delta^3(\vec{x})$, two useful 
consequences are,
\begin{equation}
\partial^2 \Bigl[a \Phi_0\Bigr] = K \delta^3(\vec{x}) - 2 a^3 H^2 \Phi_0
\qquad , \qquad \partial^2 \Bigl[ a \partial_0 \Phi_0\Bigr] = - H K a 
\delta^3(\vec{x}) \; . \label{2IDs} 
\end{equation}

The second identity of (\ref{2IDs}) can be used to perform the spatial 
integrations in the 1-loop sources induced by the 2nd and 3rd lines of 
(\ref{MSK}),
\begin{eqnarray}
\lefteqn{ \mathcal{D} \Phi_1(\eta,r) = \frac{\kappa^2 H^2}{4 \pi^2} \Bigl\{
K a \ln(a) \delta^3(\vec{x}) -3 a^3 H \partial_0 \Phi_0 - 2
\ln\Bigl(\frac{4 \mu^2 a}{H^2}\Bigr) \partial_0 \Bigl[a^2 \partial_0 \phi_0
\Bigr] \Bigr\} } \nonumber \\
& & \hspace{-0.5cm} + \frac{\kappa^2 H^3 K \partial_0}{128 \pi^3} \Biggl\{
-2 a \partial^2 \!\! \int_{\eta_i}^{\eta - r} \!\!\!\!\!\!\!\! d\eta' a' 
\ln[\mu^2 (\Delta \eta^2 \!-\! r^2)] + a (4 \nabla^2 \!-\! 9 \partial_0^2) 
\!\! \int_{\eta_i}^{\eta - r} \!\!\!\!\!\!\!\! d\eta' a' \nonumber \\
& & \hspace{-0.2cm} + 2 a (\nabla^2 \!-\! 4 \partial_0^2) \!\! 
\int_{\eta_i}^{\eta - r} \!\!\!\!\!\!\!\! d\eta' a' \ln[\frac{_1}{^4} H^2 
(\Delta \eta^2 \!-\! r^2)] \Biggr\} + \frac{\kappa^2 H^2 \nabla^2}{128 \pi^3} 
\Biggl\{ 2 a \partial^2 \!\! \int \!\! d^4x' \Theta \nonumber \\
& & \hspace{0.9cm} \times \ln[\mu^2 (\Delta \eta^2 \!-\! \Delta r^2)] 
{\partial'}^2 [a' \Phi_0(x')] - a (2 \nabla^2 \!+\! 3 \partial_0^2) \!\! 
\int \!\! d^4x' \Theta {\partial'}^2 [a' \Phi_0(x')] \nonumber \\
& & \hspace{-0.2cm} - 6 a \partial_0^2 \!\! \int \!\! d^4x' \Theta 
\ln[\frac{_1}{^4} H^2 (\Delta \eta^2 \!-\! \Delta r^2)] {\partial'}^2 
[a' \Phi_0(x')] \!+\! a^2 H^2 (16 \partial_0^2 \!+\! 6 \nabla^2) \!\! \int \!\! 
d^4x' \Theta \nonumber \\
& & \hspace{0.9cm} \times {a'}^2 \Phi_0(x') + 12 a^2 H^2 \nabla^2 \!\! 
\int \!\! d^4x' \Theta \ln[\frac{_1}{^4} H^2 (\Delta \eta^2 \!-\! \Delta r^2)] 
{a'}^2 \Phi_0(x') \Biggr\} . \qquad \label{Phi1eqn}
\end{eqnarray}
Each of the source terms on the right hand side of (\ref{Phi1eqn}) can be
evaluated, at least for late times, and then its contribution to 
$\Phi_1(\eta,r)$ can be derived by integrating against the retarded Green's
function associated with the differential operator $\mathcal{D} \equiv 
\partial^{\mu} a^2 \partial_{\mu}$,
\begin{equation}
G_{\rm ret}(x;x') = -\frac1{4\pi} \Bigl\{ \frac{\delta(\Delta \eta \!-\!
\Delta r)}{a a' \Delta r} + H^2 \theta(\Delta \eta \!-\! \Delta r)\Bigr\} 
\; . \label{Gret}
\end{equation}
However, it turns out that only the first source term in equation 
(\ref{Phi1eqn}) makes a significant contribution at late times,
\begin{eqnarray}
\lefteqn{\int \!\! d^4x' G_{\rm ret}(x;x') \!\times\! \frac{\kappa^2 H^2}{4\pi^2} 
K a' \ln(a') \delta^3(\vec{x}') } \nonumber \\
& & \hspace{4cm} = -\frac{\kappa^2 H^3 K}{32 \pi^3} \Bigl\{\ln^2\Bigl( H r \!+\! 
\frac1{a}\Bigr) - \frac{2 \ln(H r \!+\! \frac1{a})}{a H r}\Bigr\} \; . \qquad
\label{leading}
\end{eqnarray}

The easiest way to see that the other source terms in equation (\ref{Phi1eqn})
do not contribute at late times is by changing the time variable to $a$ and 
the space variable to $a H r$, and then extracting a factor of $a^4 H^2$ from
both sides of equation (\ref{Phi1eqn}). The left hand side becomes,
\begin{equation}
\mathcal{D} = a^4 H^2 \Bigl[ -a^2 \frac{\partial^2}{\partial a^2} - 4 a
\frac{\partial}{\partial a} + \frac1{a^2 H^2} \frac{\partial^2}{\partial r^2} +
\frac{2}{a^2 H^2 r} \frac{\partial}{\partial r} \Bigr] \; . \label{Dform}
\end{equation}
For an example of the right hand side, consider the second of the nonlocal 
source terms,
\begin{eqnarray}
\lefteqn{ \frac{\kappa^2 H^3 K \partial_0}{128 \pi^3} \Bigl\{ \!
a (4 \nabla^2 \!-\! 9 \partial_0^2) \!\! \int_{\eta_i}^{\eta - r} 
\!\!\!\!\!\!\!\! d\eta' a' \!\Bigr\} = \frac{\kappa^2 H^2 K \partial_0}{128 \pi^3} 
\Bigl\{ \!a (4 \nabla^2 \!-\! 9 \partial_0^2) \ln\Bigl(\!\frac1{H r \!+\! 
\frac1{a}} \!\Bigr) \! \Bigr\} ,} \\
& & \hspace{0.7cm} = a^4 H^2 \!\times\! \frac{\kappa^2 H^3 K}{128 \pi^3} \Bigl\{
-\frac{16}{a H r} \!+\! \frac{16}{a H r \!+\! 1} \!+\! \frac{3}{(a H r \!+\! 1)^2} 
\!-\! \frac{10}{(a H r \!+\! 1)^3} \Bigr\} . \qquad \label{nonlocalsource}  
\end{eqnarray}
The part of expression (\ref{nonlocalsource}) inside the curly brackets goes
like $1/(a H r)^2$ at late times, which corresponds to a late time contribution
to $\Phi_1(\eta,r)$ of the same strength according to (\ref{Dform}). The strongest 
nonlocal source goes like $\ln(a)/aHr$. Hence the leading ($a \gg 1$, $a H r \gg 1$)
form comes from (\ref{leading}),
\begin{equation}
\Phi_1(t,r) \longrightarrow -\frac{\kappa^2 H^2}{8 \pi^2} \ln(H r) 
\!\times\! \frac{H K}{4 \pi} \ln(H r) \; . \label{Phi1}
\end{equation}

\subsection{Renormalization Group Explanation}

The $h \partial \phi \partial \phi$ interaction of gravity with our scalar
is very similar to the $A \partial B \partial B$ interaction of a nonlinear
sigma model which was recently studied \cite{Miao:2021gic}. It was shown that
the leading inflationary logarithms of that model could all be explained by
combining a variant of Starobinsky's stochastic formalism 
\cite{Starobinsky:1986fx,Starobinsky:1994bd}, based on curvature-dependent 
effective potentials, with a variant of the renormalization group, based on 
curvature-dependent field strength renormalizations. The $\phi \rightarrow 
\phi + {\rm constant}$ shift symmetry of our Lagrangian (\ref{MMCSLag})
precludes there being any effective potential for $\phi$ but it does seem
possible to identify a curvature-dependent field strength renormalization
among the three 1-loop counterterms of expression (\ref{DeltaL}),
\begin{equation}
\Delta \mathcal{L} = -\frac12 \alpha_1 \square \phi \square \phi \sqrt{-g}
- \frac12 \alpha_2 R \partial_{\mu} \phi \partial_{\nu} \phi g^{\mu\nu} 
\sqrt{-g} - \frac12 \alpha_3 R \partial_0 \phi \partial_0 \phi g^{00} 
\sqrt{-g} \; .
\end{equation}
This is the $\alpha_2$ counterterm, which can be viewed as a field strength
renormalization of the original Lagrangian (\ref{MMCSLag}) with $\delta Z
= \alpha_2 R$. Just as for $A$-$B$ nonlinear sigma model of Ref. 
\cite{Miao:2021gic}, the higher derivative counterterm proportional to 
$\alpha_1$ plays no role because it cannot be viewed as a renormalization
of the bare Lagrangian. Nor can the noncovariant counterterm proportional 
to $\alpha_3$, whose existence is partially due to our use of the simple,
de Sitter breaking gauge \cite{Tsamis:1992xa,Woodard:2004ut}, and partly
to the time-ordering of interactions which seems unavoidable in the 
Schwinger-Keldysh formalism \cite{Glavan:2015ura}.

If we accept the $\alpha_2$ counterterm as a field strength renormalization,
and employ our result (\ref{alpha2}) for $\alpha_2 R$, the associated $\gamma$ 
function is,
\begin{equation}
Z = 1 + \alpha_2 R + O(\kappa^4) \qquad \Longrightarrow \qquad \gamma \equiv
\frac{\partial \ln(Z)}{\partial \ln(\mu^2)} = -\frac{\kappa^2 H^2}{8 \pi^2}
+ O(\kappa^4) \; .
\end{equation}
The exchange potential $\Phi(\eta,r)$ represents an integral of the 1PI 2-point
function, so the Callan-Symanzik equation for it should read,
\begin{equation}
\Bigl[ \frac{\partial}{\partial \ln(\mu)} - 2 \gamma \Bigr] \Phi(t,r) 
= 0 \; . \label{CSeqn}
\end{equation}
If we replace the scale parameter $\mu$ by $r$, it will be seen that equation
(\ref{CSeqn}) exactly explains the leading 1-loop logarithm from the known 
tree order result (\ref{treepot}),
\begin{equation}
\frac{\partial \Phi_1}{\partial \ln(r)} = -2 \!\times\! 
\frac{\kappa^2 H^2}{8 \pi^2} \!\times\! \Phi_0 \quad \Longrightarrow \quad 
\Phi_1(t,r) \longrightarrow -\frac{\kappa^2 H^3 K}{32 \pi^3} \ln^2(H r) \; . 
\label{logexpl}
\end{equation}

No similar explanation can be given for the late time correction 
(\ref{u1final}) to the mode function. This seems to be because $u_1(\eta,k)$ 
is not a leading logarithm effect; indeed, it vanishes at late times. 
Replacing $\mu$ by $a$, or any other time variable, is also problematic 
because most of the time dependence of the mode function comes in the form 
of $k/a H$.

\section{Conclusions}

Our long term goal is to establish the reality of large loop corrections 
from inflationary gravitons (\ref{occunum}) by purging the effective field
equations of gauge dependence. The Introduction described a procedure 
which has already been implemented on flat space background for massless,
minimally coupled scalars \cite{Miao:2017feh}, and for electromagnetism 
\cite{Katuwal:2020rkv}. We plan to generalize this procedure to de Sitter
background and, for that purpose, we have sought the simplest system which 
exhibits large graviton loop corrections before the gauge purge. Although
conformally coupled scalars are very simple, neither their mode function
nor their exchange potential shows a large correction at 1-loop order
\cite{Glavan:2020ccz}. The next simplest system is the massless, minimally
coupled scalar which we analyzed in this paper. We found that there is no
large correction to the 1-loop mode function (\ref{u1final}), but the 1-loop
exchange potential (\ref{Phi1}) does experience such a correction. Our main
conclusion is therefore that the massless, minimally coupled scalar 
(\ref{MMCSLag}) is the system we have been seeking, and its exchange 
potential is the proper thing to study.

Section 2 employed dimensional regularization to compute the fully 
renormalized 1-loop graviton contribution to the scalar self-mass 
(\ref{Mren}). Although we discovered some small mistakes in a previous 
computation \cite{Kahya:2007bc}, which do not alter the conclusion of 
that work that there are no growing secular corrections to the mode 
function, our biggest improvement is the use of
a simple representation which is not burdened with cumbersome de Sitter
invariant inverse differential operators. In section 3 we used the
result to quantum-correct the linearized, Schwinger-Keldysh effective 
field equation (\ref{phieqn}). Specializing this equation to scalar 
radiation gave relation (\ref{u1eqn}) for the 1-loop mode function,
whose asymptotic late time solution is (\ref{u1final}). We find no 
correction to the freeze-in amplitude of the mode function, but we do
find a large temporal logarithm correction to the approach to freeze-in.
Specializing (\ref{phieqn}) to the response to a point source gave 
relation (\ref{Phi1eqn}) for the 1-loop exchange potential, whose 
asymptotic late time solution is (\ref{Phi1}). We find a large spatial
logarithm correction to the exchange potential. It is significant that 
this correction derives entirely from the first of the source terms 
on the right hand side of equation (\ref{Phi1eqn}). Of course that means 
we can focus on just this term when carrying out the gauge purge, 
which is a huge simplification and justifies the effort put into this 
study.

In section 3.4 we used the renormalization group to explain the large 
logarithm in the 1-loop exchange potential (\ref{Phi1}). This is 
significant for two reasons:
\begin{itemize}
\item{It is the first time a large logarithm from inflationary gravitons
has been explained using the renormalization group; and}
\item{It ties the appearance of a large inflationary logarithm to 
the existence of the $\alpha_2$ counterterm in (\ref{DeltaL}).}
\end{itemize}
The second point is relevant to the continuing controversy over the
reality of graviton-induced logarithms because it means that the
absence of inflationary logarithms at $\ell$-loop order requires that
divergences proportional to $(G R)^{\ell} \partial_{\mu} \phi 
\partial_{\nu} \phi g^{\mu\nu} \sqrt{-g}$ must vanish, for all $\ell$,
and in the absence of any symmetry argument. That seems to strain 
credulity.

Before closing we should adumbrate the subsequent steps in our program,
its relation to the manner in which one combines Green's functions to
produce an S-matrix, and our expectations for the fate of large logarithms. 
Our aim is to purge gauge dependence from the linearized effective field 
equations of massless fields. The procedure was described in some detail 
in the Introduction, and it has been explicitly carried out on flat space 
background for gravtion corrections to the massless, minimally coupled 
scalar \cite{Miao:2017feh} and for graviton corrections to electromagnetism 
\cite{Katuwal:2020rkv}. What one does is to consider exactly the same 
Green's functions that would contribute to the scattering amplitude for the 
exchange of a massless particle between two massive particles. These Green's 
functions will include 2-point, 3-point and 4-point diagrams of the massive 
particle. Instead of going on-shell in momentum space (which would not be 
observable in cosmology) one applies a series relations derived by Donoghue 
\cite{Donoghue:1993eb,Donoghue:1994dn,Donoghue:1996mt} which reduce the 
3-point and 4-point diagrams to 2-point form, without disturbing the long-range part 
of the amplitude. The flat space forms of these relations were given in 
equations (\ref{IntID1}-\ref{IntID3}), and we propose to make the most 
straightforward generalizations to de Sitter background. The next step is 
using the propagator equation for the massless field to regard the various 
2-point diagrams as contributions to a gauge-independent 1PI 2-point function 
of the massless field, as per equation (\ref{keytrick}). We then use this 
1PI 2-point function to quantum-correct the linearized effective field 
equation of the massless field. Gauge independence can be checked by repeating
the computation in the 2-parameter family of simple, de Sitter-breaking gauges
for which the graviton propagator has been derived \cite{Glavan:2019msf}. Our
expectation is that whatever large logarithmic corrections occurred with the
gauge-dependent computation in the simplest gauge \cite{Tsamis:1992xa,
Woodard:2004ut} will persist in the gauge-independent computation but with
possibly different numerical coefficients.

\vspace{.5cm}

\centerline{\bf Acknowledgements}

DG was supported by the Czech Science Foundation (GA\v{C}R) grant 
20-28525S. SPM was supported by Taiwan MOST grants 109-2112-M-006-002 
and 110-2112-M-006-026. TP was supported by the D-ITP consortium, a 
program of the Neth\-erlands Organization for Scientific Research (NWO) 
that is funded by the Dutch Ministry of Education, Culture and Science 
(OCW). RPW was supported by NSF grant PHY-1912484 and by the Institute 
for Fundamental Theory at the University of Florida.

\end{document}